\documentstyle[twoside,11pt]{article}

\catcode`\@=11
\long\def\@makefntext#1{
\protect\noindent \hbox to 3.2pt {\hskip-.9pt  
$^{{\tenrm\@thefnmark}}$\hfil}#1\hfill}			

\def\thefootnote{\fnsymbol{footnote}}
\def\@makefnmark{\hbox to 0pt{$^{\@thefnmark}$\hss}}	
	
\def\ps@myheadings{\let\@mkboth\@gobbletwo
\def\@oddhead{\hbox{}
\rightmark\hfil\tenrm\thepage}  
\def\@oddfoot{}\def\@evenhead{\tenrm\thepage\hfil
\leftmark\hbox{}}\def\@evenfoot{}
\def\sectionmark##1{}\def\subsectionmark##1{}}

\renewcommand{\thefootnote}{\fnsymbol{footnote}}

\newcounter{sectionc}\newcounter{subsectionc}\newcounter{subsubsectionc}
\renewcommand{\section}[1] {\vspace{25pt}\addtocounter{sectionc}{1} 
\setcounter{subsectionc}{0}\setcounter{subsubsectionc}{0}\noindent 
	{\twelvebf\thesectionc.\kern0.35cm #1}\par\vspace{8pt}}
\renewcommand{\subsection}[1] {\vspace{25pt}\addtocounter{subsectionc}{1} 
	\setcounter{subsubsectionc}{0}\noindent 
	{\twelvebf\thesectionc.\thesubsectionc\kern0.35cm #1}\par 
	\vspace{8pt}}
\renewcommand{\subsubsection}[1] {\vspace{25pt}\addtocounter{subsubsectionc}{1}
	\noindent
	{\twelverm\thesectionc.\thesubsectionc.\thesubsubsectionc\kern0.35cm 
	{\kern1pt\twelveit #1}}\par\vspace{8pt}}

\newcommand{\smalllineskip}{\baselineskip=11pt}

\def\ninecirc{
\begin{picture}(0,0)
\put(4.4,1.8){\circle{7.45}}
\end{picture}}
\def\ninecopyright{\ninecirc\kern2.75pt\hbox{\eightrm c}}

\def\abstracts#1#2#3{{
	\centering{\begin{minipage}{5.0in}\tenrm\baselineskip=12pt
        \centerline{\twelvebf Abstract}
	\vspace{5pt}
	\parindent=0pc #1\par 
	\parindent=1pc #2\par
	\parindent=1pc #3
	\end{minipage}}\par}} 

\def\ARTICLES{\kern6.15cm\hbox{${\vcenter{\vbox{
	\hrule height 0.4pt width 4.562truein	    
	\hbox{\vrule width 0.4pt 		    
	height 0.6truein			    
	\raise0.565cm\hbox{\kern1pc
	\seventeenbf Articles}}
	\hrule height 0.4pt width 4.562truein}}}$}}  

\renewenvironment{thebibliography}[1]
	{\frenchspacing
	 \tenrm\baselineskip=12pt
	 \begin{list}{\arabic{enumi}.}
	{\usecounter{enumi}\setlength{\parsep}{0pt}
	 \setlength{\leftmargin 12.7pt}{\rightmargin 0pt} 
	 \setlength{\itemsep}{0pt} \settowidth
	{\labelwidth}{#1.}\sloppy}}{\end{list}}

\newcounter{itemlistc}
\newcounter{romanlistc}
\newcounter{alphlistc}
\newcounter{arabiclistc}

\newcommand{\fcaption}[1]{
        \refstepcounter{figure}
        \setbox\@tempboxa = \hbox{\footnotesize{\bf Fig.~\thefigure\phantom{00}}#1}
        \ifdim \wd\@tempboxa > 6in
           {\begin{center}
        \parbox{6in}{\footnotesize\smalllineskip{\bf Fig.~\thefigure\phantom{00}}#1}
            \end{center}}
        \else
             {\begin{center}
             {\footnotesize{\bf Fig.~\thefigure\phantom{00}}#1}
              \end{center}}
        \fi}

\newcommand{\tcaption}[1]{
        \refstepcounter{table}
        \setbox\@tempboxa = \hbox{\footnotesize\bf Table~\thetable\phantom{00}#1}
        \ifdim \wd\@tempboxa > 6in
           {\begin{center}
        \parbox{6in}{\footnotesize\smalllineskip\bf Table~\thetable\phantom{00}#1}
            \end{center}}
        \else
             {\begin{center}
             {\footnotesize\bf Table~\thetable\phantom{00}#1}
              \end{center}}
        \fi}

\def\@citex[#1]#2{\if@filesw\immediate\write\@auxout
	{\string\citation{#2}}\fi
\def\@citea{}\@cite{\@for\@citeb:=#2\do
	{\@citea\def\@citea{,}\@ifundefined
	{b@\@citeb}{{\bf ?}\@warning
	{Citation `\@citeb' on page \thepage \space undefined}}
	{\csname b@\@citeb\endcsname}}}{#1}}

\newif\if@cghi
\def\cite{\@cghitrue\@ifnextchar [{\@tempswatrue
	\@citex}{\@tempswafalse\@citex[]}}
\def\citelow{\@cghifalse\@ifnextchar [{\@tempswatrue
	\@citex}{\@tempswafalse\@citex[]}}
\def\@cite#1#2{{$\null^{#1}$\if@tempswa\typeout
	{IJCGA warning: optional citation argument 
	ignored: `#2'} \fi}}

\def\fnt#1#2{\footnotetext{\kern-.3em
	{$^{\mbox{\sevenrm #1}}$}{#2}}}

\def\runninghead#1#2{\protect\pagestyle{myheadings}
\markboth{\protect\nineit\,\,\,\,\,#1\hfill}
{\hfill\protect\nineit #2\,\,\,\,\,}}
\font\seventeenbf=cmbx10      scaled\magstep3

\font\twelverm=cmr10      scaled\magstep1
\font\twelveit=cmti10     scaled\magstep1
\font\twelvebf=cmbx10     scaled\magstep1

\font\tenrm=cmr10

\font\ninerm=cmr9
\font\nineit=cmti9

\font\eightrm=cmr8

\font\sevenrm=cmr7

\catcode`@=11					
\def\ps@plain{\let\@mkboth\@gobbletwo
     \def\@oddhead{}\def\@oddfoot{\ninerm\hfil\thepage
     \hfil}\def\@evenhead{}\let\@evenfoot\@oddfoot}

\def\ps@myheadings{\let\@mkboth\@gobbletwo	
\def\@oddhead{\hbox{}
\rightmark\hfil\ninerm\thepage}   
\def\@oddfoot{}\def\@evenhead{\ninerm\thepage\hfil
\leftmark\hbox{}}\def\@evenfoot{}
\def\sectionmark##1{}\def\subsectionmark##1{}}
\catcode`@=12

\begin{document}
\def\vsp{\par} \def\pni{\par\noindent}
\def\eg{{e.g.}\ } \def\ie{{i.e.}\ }  
\def\sg{\hbox{sign}\,}
\def\sgn{\hbox{sign}\,}
\def\sign{\hbox{sign}\,}
\def\e{\hbox{e}}
\def\exp{\hbox{exp}}
\def\ds{\displaystyle}
\def\dis{\displaystyle}
\def\q{\quad}	 \def\qq{\qquad}
\def\lan{\langle}\def\ran{\rangle}
\def\l{\left} \def\r{\right}
\def\lra{\Longleftrightarrow}
\def\arg{\hbox{\rm arg}}
\def\d{\partial}
 \def\dr{\partial r}  \def\dt{\partial t}
\def\dx{\partial x}   \def\dy{\partial y}  \def\dz{\partial z}
\def\rec#1{{1\over{#1}}}
\def\log{\hbox{\rm log}\,}
\def\erf{\hbox{\rm erf}\,}     \def\erfc{\hbox{\rm erfc}\,}
\def\FT{{\cal{F}}\,}  \def\F{{\cal{F}}\,}  
\def\LT{{\cal{L}}\,}  \def\L{{\cal{L}}\,} 
\def\NN{\hbox{\bf N}}
\def\RR{\hbox{\bf R}}
\def\CC{\hbox{\bf C}}
\def\ZZ{\hbox{\bf Z}}
\def\II{\hbox{\bf I}}


\runninghead{Derivation of the fractional diffusion equation}
{Derivation of the fractional diffusion equation}

\renewcommand{\thefootnote}{\fnsymbol{footnote}}      

\thispagestyle{plain}
\setcounter{page}{1}


\vspace{6pc}
\leftline{\phantom{\ARTICLES}\hfill}

\vspace{3pc}
\leftline{\hskip-0.1cm\vbox{\hrule width6.99truein height0.15cm}\hfill}

\vspace{2pc}
\centerline{\seventeenbf REVISITING THE DERIVATION OF THE}
\baselineskip=20pt
\centerline{\seventeenbf FRACTIONAL DIFFUSION EQUATION}
\vspace{0.27truein}
\centerline{ENRICO SCALAS}
\baselineskip=12.5pt
\centerline{\it Dipartimento di Scienze e Tecnologie Avanzate,}
\centerline{\it Universit\`a del Piemonte Orientale}
\centerline{\it Corso Borsalino 54, Alessandria, I--15100, Italy}
\vspace{0.08truein}
\centerline{RUDOLF GORENFLO}
\baselineskip=12.5pt
\centerline{\it Erstes Mathematisches Institut, Freie Universit\"at Berlin}
\centerline{\it Arnimallee 3, Berlin, D--14195, Germany}
\vspace{0.08truein}
\centerline{FRANCESCO MAINARDI}
\baselineskip=12.5pt
\centerline{\it Dipartimento di Fisica, Universit\`a di Bologna e INFN Sezione
di Bologna}
\centerline{\it Via Irnerio 46, Bologna, I--40126, Italy}
\vspace{0.08truein}
\centerline{MARCO RABERTO}
\baselineskip=12.5pt
\centerline{\it Dipartimento di Ingegneria Biofisica ed Elettronica, 
Universit\`a di Genova}
\centerline{\it Via all'Opera Pia 11a, Genova, I--16145, Italy}
\vspace{0.36truein}
\abstracts{The fractional diffusion equation is derived from
the master equation of continuous time random walks (CTRWs) via a
straightforward application of the Gnedenko-Kolmogorov 
limit theorem. The Cauchy problem for the fractional diffusion equation
is solved in various important and general cases. The meaning
of the proper diffusion limit for CTRWs is discussed.}{}{}

\vspace{0.78truein}
\baselineskip=14.5pt
\section{INTRODUCTION}

\noindent This paper provides a short, but self-contained, introduction to fractional
diffusion. The readers will find the basic ideas behind the derivation
of the fractional diffusion equation starting 
from continuous-time random walks.
We have included formulae for the solution of the Cauchy
problem which can be numerically implemented and used for applications.
Special care has been used to avoid unessential mathematical technicalities.
Even if far 

\newpage

\noindent from exhaustive, 
the bibliography should give a sufficient number of 
entry points for further reading.
The following sections are based on a series 
of papers about the application of
fractional calculus to finance\cite{SGM 00,MRGS 00,RSGM 01,GMSR 01}.

It is our hope that theoretical and experimental condensed matter 
physicists will find this work useful. 

The paper is divided as follows. In Section 2, we outline the theory
leading to the time-fractional master equation.
In Section 3, the transition to the
space-time fractional diffusion equation is discussed. Section
4 is devoted to the solutions of the Cauchy problem for
the fractional diffusion equation. The main results are 
briefly summarized and discussed in Section 5.
In Appendix A and Appendix B, we introduce
the definitions of fractional derivatives
in time and space, respectively, entering the
fractional diffusion equation.

\section{STEP ONE: TRANSITION TO THE TIME-FRACTIONAL MASTER EQUATION}
\noindent
Let $x$ be the position of a diffusing particle in one dimension.
Let us assume that both jumps $\xi_i = x(t_i) - x(t_{i-1})$ and waiting times
between two consecutive jumps $\tau_i =t_i - t_{i-1}$ are i.i.d.
random variables described by two probability density functions: $w(\xi)$ and
$\psi(\tau)$.
According to the model of continuous-time
random-walk (CTRW), introduced by Montroll and
Weiss\cite{Montroll 65,Shlesinger 96},
the evolution equation for $p(x,t)$, the probability
of finding the random walker at position $x$ at time instant $t\,, $
can be written as follows, assuming the initial condition $p(x,0) = \delta (x) $
(\ie the walker is initially at the origin $x=0)$\cite{MRGS 00},
$$   p(x,t) =  \delta (x)\, \Psi(t) +
   \int_0^t   \psi(t-t') \, \l[
 \int_{-\infty}^{+\infty}  w(x-x')\, p(x',t')\, dx'\r]\,dt'
 \,, \eqno(2.1) $$
where
  $$ \Psi(t) =\int_t ^\infty \psi(t')\, dt'
  = 1- \int_0^t \psi(t')\, dt'\,, \q
 \psi(t) = - {d \over dt} \Psi(t)\,. \eqno(2.2)$$
The {\it master equation}
of the CTRW can be also derived in the Fourier-Laplace domain.

The integral $\int_0^\tau  \psi(t')\, dt'\,$ represents the
probability  that at
least one  step is taken at some instant in the interval $[0,\tau) $.
Thus, $\Psi(\tau )\,$ is the probability  that the diffusing quantity
$x$ does not change value during the time interval of duration $\tau $
after a jump.

In a paper by Mainardi et al.\cite{MRGS 00}, an alternative form 
of Eq. (2.2) was presented
in terms of a convolution between the first time derivative of $p(x,t)$
and a suitable kernel. The resulting equation  can be interpreted as
an {\it evolution} equation
of generalized {\it Fokker-Planck-Kolmogorov} kind.
It reads:
$$
\int_{0}^{t} \Phi(t-t') \,{\partial \over \partial t'} p(x,t')\, dt' =
- p(x,t) +
\int_{-\infty}^{+\infty} w(x-x') \,p(x',t)\,dx'\,,
\eqno(2.3)
$$
where the ``auxiliary'' function
$\Phi(t)\,$
is such that
$\Psi(t) = \int_0^t\Phi (t-t') \, \psi(t' )\,dt' \,.$
Eq. (2.3) can be obtained by Fourier-Laplace transforming
Eq. (2.1) and by suitable assumptions on the Laplace transform
of the function $\Phi(t)$.

In general, a CTRW is a non-Markovian process.
A CTRW becomes Markovian  if (and only if)  the above
memory function is proportional to a delta function so
that $\Psi(t)$ and $\psi(t)$ differ only by a multiplying positive constant.
By an appropriate choice of the unit of time, we can write
 $ \Phi(t) =  \delta (t)\,,\; t\ge 0\,.$
In this case,
Eq. (2.3) becomes:

\newpage

$$  {\d \over \d t} p(x,t) = - p(x,t)  +
   \int_{-\infty}^{+\infty}   w(x-x')\,p(x',t)\,dx'\,,
     \q p(x,0) = \delta(x)\,. \eqno(2.4) $$

\noindent Up to a change of the unit of time,
this is the most general {\it master equation} for a {\it Markovian}
$CTRW$; Saichev \& Zaslavsky call it the 
{\it Kolmogorov-Feller equation}\cite{Saichev 97}.

Eq. (2.3) allows a natural characterization of a peculiar
class of non-Markovian processes, where the memory function,
$\Phi(t)\,$ has power-law time decay.
Within this class, an interesting choice is the following:
$$
    \Phi(t)   = {t^{-\beta}\over  \Gamma(1-\beta)}\,,  \q t\ge 0\,,
 \q 0<\beta <1\,. \eqno(2.5)$$
In this case, $\Phi(t)$ is a weakly singular function
that, in the limit $\beta \to 1\,, $
reduces to  $\Phi(t) =	\delta (t)\,, $
according to  the formal representation of the Dirac generalized function,
$\delta(t) = t^{-1}/\Gamma(0)\,, \; t \ge 0$\cite{Gelfand 64}.
As a consequence of the choice (2.5)\cite{MRGS  00},
Eq.  (2.3) can be written as:
$$ {\d^\beta   \over \d t^\beta } p(x,t) =
     -	p(x,t) +   \int_{-\infty}^{+\infty} w(x-x')\,
   p(x',t) \, dx'\,, \q p(x,0) = \delta(x)\,,
\eqno(2.6) $$
where ${\d^\beta  / \d t^\beta } $
is the pseudo-differential operator
explicitly defined in the Appendix A, that we usually call the
{\it Caputo} fractional derivative of order $\beta  \,. $
Eq. (2.6) is a {time-fractional
generalization} of Eq. (2.4) and can be
called {\it time-fractional Kolmogorov-Feller equation}.

Our choice  for $\Phi(t)$  implies peculiar forms
for the functions $\Psi(t)$ and $\psi(t)$  
generalizing the exponential behaviour of the waiting time
density in the Markovian case.
In fact, we have for $t\ge 0$\cite{MRGS 00}:
$$\Psi(t) =  E_\beta (-t^\beta)
\,,\q \psi(t) =  
	    -	{d \over dt}  E_\beta (-t^\beta)
 \,, \q 0<\beta < 1\,,\eqno (2.7)$$
where $E_{\beta}$ denotes an entire transcendental function, known as
the Mittag-Leffler function of order $\beta\,,$
defined in the complex plane by the power series
$$ E_\beta (z) :=
    \sum_{n=0}^{\infty}\,
   {z^{n}\over\Gamma(\beta\,n+1)}\,, \q \beta >0\,, \q z \in \CC\,.
 \eqno	(2.8)$$
Detailed information on the Mittag-Leffler-type functions
is available in the literature\cite{Erdelyi HTF,GorMai 97,MaiGor 00}.

From the properties of the Mittag-Leffler function, it can be shown that
the corresponding
survival probability and waiting-time $pdf$
interpolate between a stretched exponential, for small waiting
times, and a power-law decay, for large waiting times.
Such a behaviour has been observed in
Mainardi et al.\cite{MRGS 00} and in Raberto et al.\cite{RSGM 01}, where
the function $\Psi(\tau )$ has been estimated from empirical
financial data.

As a final remark, it is important to notice that different
choices of the kernel $\Phi(t)$ in eq. (2.3) are possible,
leading to different properties of the waiting-time pdf
and different generalized Kolmogorov-Feller evolution equations
for $p(x,t)$.

\newpage

\section{STEP TWO: THE DIFFUSION LIMIT}
\noindent
In the physical literature, 
many authors have discussed the connection between continuous--time random 
walks and diffusion equations of fractional order, with different degrees
of detail\cite{RomanAlemany 94}$^{-}$\cite{Uchaikin 00}.  
A sound proof of the equivalence between fractional diffusion and CTRW
has been given by Hilfer \& Anton\cite{HilferAnton 95}.
However, in order to perform the transition to the diffusion limit, we
shall use a
different approach. We shall start from Eqs. (2.4)
and (2.6), and pass through their Fourier-Laplace counterparts.
The stochastic process 
whose probability density evolves according to those equations 
is a random walk originating
from a sequence  of jumps,
each jump being a sample of a real random variable $Y$. 
During the time interval
$t_n \le t < t_{n+1}\,, $ the particle
position is $Y_1 +Y_2 + \dots Y_n\,.$  The $Y_k$ are $i.i.d.$ random variables
all described, as $Y $, by the $pdf$ $w(x)\,$.
Let us denote by $\widehat w(\kappa )\,$ the characteristic
function corresponding to the probability density $w(x)$.

Let us specify some conditions on the $pdf$ $w(x)$. The
requirement is that, if $\alpha =2\,:$
$$
\sigma ^2 = \int_{-\infty}^{+\infty}  x^2\, w(x)\, dx <\infty\,,
 \eqno(3.1)$$
whereas, if  $0<\alpha<2\,:$
$$  w(x) = \l( b +\epsilon (|x|) \r)|x|^{-(\alpha +1)}\,,
 \q b>0\,,\q \epsilon(|x|) \to 0 \;\hbox{as} \; |x| \to \infty\,.
  \eqno(3.2)$$
In Eq. (3.2), $b>0$ and $\epsilon(|x|)$ is bounded and
$ O \l(|x|^{-\eta}\r)\,$ with $\eta >0$ as $	|x|\to	 \infty\,.$
Let us furthermore recall the necessary requirements
$ w(x) \ge 0 \,,$
and	the normalization condition
$	 \int_{-\infty}^{+\infty} w(x) \, dx =1\,. $
\vsp
Let us now consider a sequence of random process $pdf'$s $p_h(x,t)$
describing scaled jumps of size  $h\, Y_k\, $ instead
of $Y_k\,,$ with a speed increase of the process by a factor
({\it the scaling factor})
$\mu ^{-1/\beta }\, h^{-\alpha /\beta }\,, $
where $\mu  $ must satisfy some conditions which will be 
specified later.
The $pdf$ of the jump size is $w_h(x) = w(x/h)/h\,,$ so that its
characteristic function is
$\widehat {w_h}(\kappa ) = \widehat w (\kappa h)\,. $
For $0<\alpha \le 2\, $ and $\, 0<\beta \le 1\, $,
Eq. (2.6) (including (2.4) in the special case $\beta =1$)
is replaced by the sequence of equations
$$\mu h^\alpha\,{\d^\beta   \over \d t^\beta } p_h(x,t) =
     -	p_h(x,t) +   \int_{-\infty}^{+\infty} \! w_h  (x-x')\,
   p_h(x',t) \, dx'\,.\eqno(3.3) $$
By Fourier-Laplace transforming and by recalling
the Laplace transform of the Caputo time-fractional derivative,
defined by Eq. (A.2),
we have
$$ \mu h^\alpha\,\l\{
 s^\beta  \, \widehat{\widetilde  p}_h(\kappa ,s) - s^{\beta -1}\r\}
    = \l[\widehat w_h  (\kappa)-1\r]\,
   \widehat{\widetilde	p}_h(\kappa ,s) \,.
\eqno(3.4) $$

We shall now present arguments
based on the {\it classical central limit theorem}
or on the {\it Gnedenko limit theorem},
(see the book by Gnedenko \& Kolmogorov\cite{GnedenkoKolmogorov 54})
both expressed in terms of the characteristic functions.
The Gnedenko limit theorem is a suitable
generalization of the classical central limit theorem
for space $pdf'$s with infinite variance, decaying according
to condition (3.2).

\vsp
The transition to the {\it diffusion limit} is based on the following
Lemma introduced by Gorenflo\cite{GorMai CHEMNITZ01}:

\newpage

\pni 
{\it With the scaling parameter}
$$ \mu = \cases{
     {\ds {\sigma ^2\over 2}}\,,  \;& if $\q \alpha =2 \,,$ \cr
    {\ds {b\,\pi \over \Gamma(\alpha +1)\,\sin(\alpha \pi/2)}}\,,\;
    & if $\q 0<\alpha <2 \,,$ \cr}
\eqno(3.5)$$
{\it we have the relation}
$$ \lim_{h\to 0} { \widehat w(\kappa h) -1\over \mu \, h^\alpha  }
    = -|\kappa |^\alpha \,, \q 0<\alpha\le 2\,,
  \q \kappa \in\RR\,. \eqno(3.6)$$

Now, it is possible to set
$$ \rho _h(\kappa )=  {\widehat w(\kappa h)-1 \over \mu h^\alpha}  \,,
   \eqno(3.7) $$

\noindent and the sequence of equations (3.4) reads
$$  s^\beta  \, \widehat{\widetilde {p_h}}(\kappa ,s) - s^{\beta -1}
    = \rho _h  (\kappa)\,
   \widehat{\widetilde{p_h}}(\kappa ,s) \,.
       \eqno(3.8)$$
Then, passing to the limit $h\to 0$, thanks to (3.6), we get:
$$  s^\beta  \, \widehat{\widetilde{p_0}}(\kappa ,s) - s^{\beta -1}
    = -|\kappa|^\alpha \,
   \widehat{\widetilde{p_0}}(\kappa ,s) \,,
 \q 0<\alpha \le 2\,,\q 0<\beta \le 1\,.      \eqno(3.9)$$
By inversion and using the Fourier transform
of the Riesz space-fractional derivative, defined in Eq. (B.2),
we finally obtain the equation:
$$ {\d^\beta   \over \d t^\beta } p_0(x,t) =
      {\d^\alpha \over \d |x|^\alpha} p_0(x,t)
   \,, \q p_0(x,0) = \delta(x)\,,
     \eqno(3.10) $$
which is a space-time fractional
diffusion equation. In the limiting cases
$\beta =1$ and $\alpha =2$, Eq. (3.10) reduces
to the standard diffusion equation.

We have presented a formally correct transition to the diffusion limit
starting from the general master equation of the CTRW,
namely Eq. (2.1) or Eq. (2.3).
By invoking the continuity theorem of probability theory,
see \eg the book by Lukacs\cite{Lukacs 60}, we can see that the
random variable whose density is $p_h(x,t)$ converges
in distribution ("weakly" or "in law") to the random variable
with density $p_0(x,t)\,. $

Solving (3.8) for $\widehat{\widetilde{p_h}}(\kappa ,s)\,, $
and (3.9) for $\widehat{\widetilde{p_0}}(\kappa ,s)\,, $
gives:
$$    \widehat{\widetilde  {p_h}}(\kappa ,s) =
       {s^{\beta -1}\over s^\beta - \rho _h(\kappa )}	\,,
\q    \widehat{\widetilde  {p_0}}(\kappa ,s) =
       {s^{\beta -1}\over s^\beta + |\kappa|^\alpha }
\,,		    \eqno(3.11) $$
which yields:
$$ \widehat {p_h}(\kappa ,t) =
    E_\beta \l( \rho _h(\kappa) t^\beta \r)\,,
 \q \widehat {p_0}(\kappa ,t) =
    E_\beta \l( -|\kappa|^\alpha  t^\beta \r)\,.
   \eqno(3.12)$$
By (3.6) $\rho _h(\kappa ) \to -|\kappa |^\alpha $ as $h\to 0\,, $
hence
  $$ p_h(x ,t) \to  p_0(x ,t) \,, \q \hbox{for} \q
     t>0\,, \q h \to 0\,.\eqno(3.13) $$

\newpage

\section{SOLUTIONS AND THEIR SCALING PROPERTIES}
\noindent
For the determination of the fundamental solutions of Eq. (3.12) in
the general case $\{0<\alpha \le 2\,,$ $\,0<\beta \le 1\}$
the reader can consult Gorenflo et al.\cite{GoIsLu 00} and Mainardi et
al.\cite {LuMaPa 01}.
We also refer to the above references for the
particular cases $\{0<\alpha \le 2\,,$ $\,\beta = 1\}$
and $\{\alpha = 2\,,$ $\,0< \beta \le 1\}$, already
dealt with in the literature.

For parameters in the interval
$0< \alpha \le 2$, and $0< \beta \le 1$,
the Cauchy problem in Eq. (3.12) can be solved
by means of the Fourier-Laplace transform method.

The solution (Green function) turns out to be:
$$
p_{0}(x,t) = \frac{1}{t^{\beta/\alpha}} W_{\alpha,\beta} \left ( \frac{x}{t^{\beta/\alpha}}
\right ).
\eqno(4.1)
$$
The function $W_{\alpha,\beta} (u)$ is the Fourier transform of a
Mittag-Leffler function:
$$
W_{\alpha,\beta} (u) = \frac{1}{2 \pi} \int_{-\infty}^{+\infty}
\e^{-iqu} E_{\beta} (-|q|^{\alpha}) dq.
\eqno(4.2)
$$

Indeed, $E_{\beta}$ is the Mittag-Leffler function of order $\beta $ and argument
$z= - |q|^{\alpha}$

In the limiting case $0 < \alpha < 2$ and $\beta =1$, the solution is:
$$
p_{0}(x,t) = \frac{1}{t^{1/\alpha}} L_{\alpha} \left ( \frac{x}{t^{1/\alpha}}
\right ),
\eqno(4.3)
$$
where $L_{\alpha} (u)$
is the L\'evy standardized probability density function:
$$
L_{\alpha} (u) = \frac{1}{2 \pi} \int_{-\infty}^{+\infty}
\e^{-iqu-|q|^{\alpha}} dq,
\eqno(4.4)
$$
whereas, in the case $\alpha=2$, $0 < \beta < 1$\cite{Mainardi 96},
we get
$$ p_{0} (x,t) =
{1\over 2} t^{-\beta /2}\, M_{\beta/2} \left (
\frac{x}{t^{\beta /2}} \right )\,,
\eqno(4.5)$$
where
$M_{\beta/2}$ denotes  the $M$ function of Wright type
of order $\beta/2\,.$

Remarkably, a composition rule holds true, and it can be shown that
the Green function for the space-time fractional diffusion equation of order 
$\alpha$ and $\beta$ can be written in terms of the Green function
for the space-fractional diffusion equation of order $\alpha$ and
the Green function for the time-fractional diffusion equation
of order $2 \beta$\cite{LuMaPa 01}:

$$  p_{0}(x ,t) =
   t^{ -\beta } \,
  \int_0^\infty r^{-1/\alpha}\,  {L}_{\alpha}\l(x/r^{1/\alpha }\r) \,
      {M}_{\beta}\l(r/t^{\beta}\r)\, dr \,. \eqno (4.6)$$

Finally, as written before, in the case $\alpha = 2$, $\beta =1$,
Eq. (3.12) reduces to the standard diffusion equation, and the Cauchy
problem is solved by:

$$ p_{0}(x,t) = t^{-1/2}\, {1\over 2\sqrt{\pi}} \,\exp (-x^2/(4t)) =
   t^{-1/2}\, G \left( \frac{x}{t^{1/2}} \right )\,,  \eqno(4.7)$$
where $G(x)$ denotes the Gaussian $pdf$ 
$$ 
 G(x) = {1\over 2\sqrt{\pi}}\,\exp (-x^2/4)\,.\eqno(4.8)$$

\newpage

\section{SUMMARY AND DISCUSSION}
\noindent
Applications of fractional diffusion equations have 
been recently reviewed by Uchaikin \& Zolotarev\cite{Uchaikin 99}
and by Metzler \& Klafter\cite{MetzlerKlafter 00}. 
After that, other contributions appeared on this 
issue, among which we quote the paper of Zaslavsky
in the book edited by Hilfer\cite{Zaslavsky 00},
the papers by Meerschaert er al.\cite{Meer 01} 
and by Paradisi et al.\cite{Paradisi 01} and the letter
by West and Nonnenmacher\cite{West 01}.

In this paper, a scaling method has been discussed to get
the transition to the diffusion limit
in a correct way, starting from the CTRW master equation 
describing the time evolution of a stochastic 
process. Moreover, the solutions of the Cauchy problem for the 
fractional diffusion equation have been listed for the various relevant
values of the fractional derivative orders $\alpha$ and $\beta$.

Various formulae which can be useful
for applications have been presented.
In principle, given a diffusing quantity, 
the waiting--time density, the
jump density, and the probability of finding
the random walker in position $x$ at time $t$ are all 
quantities which can be empirically determined. 
Therefore, many
relationships presented above can be corroborated or 
falsified in specific contexts.

As a further remark, it may be useful to add some comments on the meaning of
the diffusion limit taken in section 3.

The factor $\mu  h^\alpha$ can be viewed as causing the jump process to
run faster and faster (the waiting times becoming shorter and shorter)
as $h$ becomes smaller and smaller. Replacing the density $w(x)$
by the density $w_h(x) = w(x/h)/h\,,$ and, accordingly, the jumps
$Y$ by $hY\,,$ means that the jump size becomes smaller and smaller
as the scaling length $h$ tends to zero.

An alternative interpretation is that we look at 
the same process with a discrete number of jumps occurring after finite times,
from far away and after long time,
so that spatial distances and time intervals of normal size
appear very small, being
$x$ replaced by $x/h\,,$ $t$ replaced by
$t/(\mu ^{1/\beta} \,h^{\alpha/\beta})\,. $

\newpage

\noindent{\bf APPENDIX A: THE CAPUTO TIME-FRACTIONAL DERIVATIVE}\\

\noindent
For readers' convenience, here, we present an
introduction to the {\it Caputo} fractional derivative starting from its
representation in the Laplace  domain and pointing out its difference
from the standard {\it Riemann-Liouville} fractional derivative.
In so doing we avoid
the subtleties lying in the inversion of fractional integrals.

If $f(t)$ is a (sufficiently well-behaved) function  with Laplace
transform
$  \; {\L} \l\{ f(t);s\r\}= \widetilde f(s)
 = \int_0^{\infty} \e^{\ds \, -st}\, f(t)\, dt\,,
$
we have
$$  {\L} \l\{ {d^\beta \over d t^\beta} f(t);s\r\} =
    s^\beta \, \widetilde f(s) - s^{\beta-1}\, f(0^+)\,, \q 0<\beta <1\,,
 \eqno(A.1) $$
if we define
$$ {d^\beta \over d t^\beta} \,f(t) :=
      {1 \over \Gamma(1-\beta )}\,\int_0^t
 {df(\tau )\over d\tau }\, {d\tau \over (t-\tau )^{\beta}} \,.
 \eqno(A.2) $$
We can also write
$$ {d^\beta \over d t^\beta} f(t)=
  {1 \over \Gamma(1-\beta)}\,{d  \over d t} \l\{
   \int_0^t
   [f(\tau )- f(0^+)]\,
  {d \tau \over (t-\tau )^{\beta}} \r\}\,,
\eqno(A.3)$$	
   $$ {d^\beta \over d t^\beta} f(t)=
  {1 \over \Gamma(1-\beta )}\,{d \over d t} \l\{
    \int_0^t
  {f(\tau )\over (t-\tau )^{\beta}} \,d \tau \r\}
  - {t^{-\beta }\over \Gamma(1-\beta)}\, f(0^+) \,. \eqno(A.4)$$

Eqs. (A.1-4) can be extended to any non integer $\beta >1\,, $
(see \eg the survey by Gorenflo \& Mainardi\cite{GorMai 97}).
We refer to the fractional derivative defined by (A.2)	as
the {\it Caputo} fractional derivative, as it was
used by Caputo
for modelling  dissipation effects in 
{\it linear viscoelasticity}
in the late sixties\cite{Caputo 67,Caputo 69,CaputoMaina 71}.

This definition
differs from the usual one named after
Riemann  and Liouville, given by the first term in
the R.H.S. of (A.4), and
defined \eg in the treatise on Fractional Calculus by Samko, Kilbas \&
Marichev\cite{SKM 93}.

Gorenflo \& Mainardi\cite{GorMai 97} and Podlubny\cite{Podlubny 99}
have pointed out the usefulness of the
Caputo fractional derivative
in the treatment of differential equations of fractional
order for {\it physical applications}.
In fact, in physical problems, the initial conditions are usually
expressed in terms of a given number of boundary values assumed by the
field variable and its derivatives of integer order,
despite the fact that
the governing evolution equation may be a generic integro-differential
equation and therefore, in particular, a  fractional differential
equation.


\newpage

\noindent{\bf APPENDIX B: THE RIESZ SPACE-FRACTIONAL DERIVATIVE}\\

\noindent
If $f(x)$ is a (sufficiently well-behaved) function  with Fourier
transform
$$ \F \l\{ f(x);\kappa \r\} = \hat f(\kappa)
  = \int_{-\infty}^{+\infty} \e^{\,\ds i\kappa x}\,f(x)\, dx\,,
  \q \kappa \in \RR\,, $$
we have
$$ \F \l\{ {d^\alpha\over d|x|^\alpha} f(x);\kappa \r\} =
    -|\kappa|^\alpha  \, \hat f(\kappa) \,, \q 0<\alpha  <2\,,
\eqno(B.1)$$
if we define
$$ {d^\alpha  \over d |x|^\alpha} f(x)	=
 \Gamma(1+\alpha ) \,
 {\sin \,(\alpha \pi/2) \over \pi }\,
 \int_0^\infty
 {f(x+\xi)- 2f(x) + f(x-\xi) \over {\xi}^{1+\alpha}}\, d \xi
 \,.  \eqno(B.2)$$
The fractional derivative defined by (B.2) can be called
{\it Riesz fractional derivative}, as it is obtained from the
inversion of the fractional integral originally introduced by
Marcel Riesz, known as the {\it Riesz potential}\cite{SKM 93}.	
The representation (B.2)\cite{GorMai CHEMNITZ01},  
is more explicit and convenient than
others found in the literature\cite{Saichev 97,SKM 93}.
It is based on a suitable regularization of a hyper-singular integral.

For $\alpha =2$, the Riesz derivative reduces to the standard
derivative of order 2, as $-|\kappa|^2 = - \kappa ^2\,. $

For $\alpha =1$, the Riesz derivative  is related to the
Hilbert transform, 
resulting in the formula
$$  {d	\over d |x|} f(x)  =
  -{1 \over \pi }\,{d \over dx}\, \int_{-\infty}^{+\infty}
   {f(\xi)\over x-\xi} \,d \xi \,.\eqno(B.3) $$

We note, by writing
$- |\kappa| ^\alpha = - (\kappa^2)^{\alpha /2},	 $
 that the Riesz
derivative  of order $\alpha$ can be interpreted as the opposite of the
$\alpha/2$ power of the (positive definite) operator
$ - D^2 = -{d^2\over dx^2}\,,$ namely
$$ {d^\alpha  \over d|x|^\alpha} =
	- \l ( - {d^2\over dx^2}  \r) ^{\alpha /2}\,. \eqno(B.4)
$$
The notation used above is due to Saichev \& Zaslavsky\cite{Saichev 97}.
 A different notation which takes into account asymmetries was 
 used by
 Gorenflo \&  Mainardi\cite{GorMai FCAA98,GorMai ZAA99}.


\end{document}